# Symmetry constraints on spin order transfer in parahydrogen-induced polarization (PHIP)

Andrey N. Pravdivtsev[1,*], Danila A. Barskiy[2,*], Jan-Bernd Hövener[1] and Igor V. Koptyug[3]

[1] Section Biomedical Imaging, Molecular Imaging North Competence Center (MOIN CC), Department of Radiology and Neuroradiology, University Medical Center Schleswig-Holstein (UKSH), Kiel University, Am Botanischen Garten 14, 24118 Kiel, Germany
[2] Helmholtz-Institut Mainz, GSI Helmholtzzentrum für Schwerionenforschung, 55128 Mainz, Germany, Johannes Gutenberg Universität Mainz, 55099 Mainz, Germany
[3] International Tomography Center, SB RAS, 3A Institutskaya st., 630090 Novosibirsk, Russia
* Correspondence: andrey.pravdivtsev@rad.uni-kiel.de and dbarskiy@uni-mainz.de

**Abstract:**

It is well known that the association of parahydrogen ($pH_2$) with an unsaturated molecule or a transient metalorganic complex can enhance the intensity of NMR signals; the effect is known as parahydrogen-induced polarization (PHIP). During the last decades, numerous methods were proposed for converting $pH_2$-derived nuclear spin order to the observable magnetization of protons or other nuclei of interest, usually $^{13}C$ or $^{15}N$. Here, we analyze the constraints imposed by the topological symmetry of the spin systems on the amplitude of transferred polarization. In asymmetric systems, heteronuclei can be polarized to 100%. However, the amplitude drops to 75% in $A_2BX$ systems and further to 50% in $A_3B_2X$ systems. The latter case is of primary importance for biological applications of PHIP using sidearm hydrogenation (PHIP-SAH). If the polarization is transferred to the same type of nuclei, i.e. $^1H$, symmetry constraints impose significant boundaries on the spin-order distribution. For AB, $A_2B$, $A_3B$, $A_2B_2$, AA'(AA') systems, the maximum average polarization for each spin is 100%, 50%, 33.3%, 25%, and 0, respectively, when A and B (or A') came from $pH_2$ We also discuss the effect of dipole-dipole induced $pH_2$ spin-order distribution in heterogeneous catalysis or nematic liquid crystals. Practical examples from the literature illustrate our theoretical analysis.

## 1. Introduction

Parahydrogen-induced polarization (PHIP) is a cost-efficient method to polarize nuclear spins[1]. PHIP exploits the symmetry of molecular dihydrogen that exists as two nuclear spin isomers: parahydrogen ($pH_2$) and orthohydrogen ($oH_2$). The nuclear spin state of $pH_2$ is the singlet state, $|S\rangle = \frac{|\alpha\beta\rangle - |\beta\alpha\rangle}{\sqrt{2}}$, which is assymetric under exchange of the nuclear spins. The total wave functions of the $H_2$ nuclei is antisymmetric under exchange of two nuclei (two fermions), so that the quantum numbers of the rotational states take even values [2]. $oH_2$ is represented by three nuclear spin states, $|T_0\rangle = \frac{|\alpha\beta\rangle + |\beta\alpha\rangle}{\sqrt{2}}$, $|T_+\rangle = |\alpha\alpha\rangle$, $|T_-\rangle = |\beta\beta\rangle$. These three states are symmetric, hence necessitating odd rotational quantum numbers[2]. This selection is dictated by the generalized Pauli principle, which states that the total wave function of two protons (two fermions with spin-½) is antisymmetric upon permutation.[2] Note, however that hydrogen consisting of a proton and an electron (i.e. two fermions) is a is a boson, hence total wave function of $H_2$ is symmetric under exchange of two atoms (discussed more below). The gap between the lowest two rotational energy levels, i.e. $pH_2$ and $oH_2$, is significant (170.5 K), so that 50% $pH_2$ can be obtained by cooling $H_2$ to liquid nitrogen temperatures [3] or even higher enrichment of 99% with a two-stage cryo-systems operating at 20 K [4,5].

The density matrix for an ensemble of molecules containing N spin-½ nuclei (with spins A and B originating from pH$_2$ molecule) can be written as follows:

$$\hat{\rho}_S^{A,B} = \frac{\hat{1}^N}{2^N} - \frac{1}{2^{N-2}}(\hat{\mathbf{I}}^A \cdot \hat{\mathbf{I}}^B). \quad (1)$$

Here, $\hat{1}^N$ is the identity matrix, i.e., a $\{2^N \times 2^N\}$ matrix with ones on the diagonal. The individual spin operators $\hat{I}_k^{A,B}$ in the dot product, $(\hat{\mathbf{I}}^A \cdot \hat{\mathbf{I}}^B) = \hat{I}_X^A \hat{I}_X^B + \hat{I}_Y^A \hat{I}_Y^B + \hat{I}_Z^A \hat{I}_Z^B$, are obtained using the Kronecker (direct) product $\otimes$ of the corresponding Pauli matrices $\hat{s}_k$ (with k = X, Y or Z) with the 2 × 2 identity matrix $\hat{1}^1$. Here, the numbering of the spins in the molecule is important. For example, for the first spin, the operator is constructed as

$$\hat{I}_k^1 = \frac{1}{2}\hat{s}_k \otimes \hat{1}^1 \ldots \otimes \hat{1}^1. \quad (2)$$

There are two primary variants of PHIP: (a) hydrogenative PHIP, such as PASADENA (parahydrogen and synthesis allow dramatically enhanced nuclear alignment [6]) and ALTADENA (adiabatic longitudinal transport after dissociation engenders net alignment [7]), and (b) non-hydrogenative PHIP, or SABRE (signal amplification by reversible exchange [8]), where pH$_2$ and substrate interact via a reversible exchange at a catalyst. Both methods have found applications at high (~ T) [9], low ~ 1 mT [10], ultra-low ~ 1 μT [11,12] and zero fields [13]. To limit the scope of this paper, however, we focus our discussion on hydrogenative PHIP at high magnetic fields only. It should be noted that a similar analysis for four spin-½ SABRE system was recently performed [14].

For hydrogenative PHIP, the spin state of the molecule after pH$_2$ addition strongly depends on the coupling regime. Two spins $\hat{\mathbf{I}}^A$ and $\hat{\mathbf{I}}^B$ are considered strongly coupled when the difference of their Larmor precession frequencies, $\delta\nu_{AB} = |\nu_A^0 - \nu_B^0|$, is much smaller than their mutual indirect spin-spin coupling $J_{AB}$, i.e., $\delta\nu_{AB} \ll |J_{AB}|$. In the opposite case, the spins are weakly coupled [15]. The frequency $\nu_{A,B}^0 = \gamma_{A,B} B_0 (1 + \delta_{A,B})/2\pi$ of spin $\hat{\mathbf{I}}^A$ or $\hat{\mathbf{I}}^B$ depends on the strength of magnetic field $B_0$, chemical shift $\delta_{A,B}$ and magnetogyric ratio $\gamma_{A,B}$.

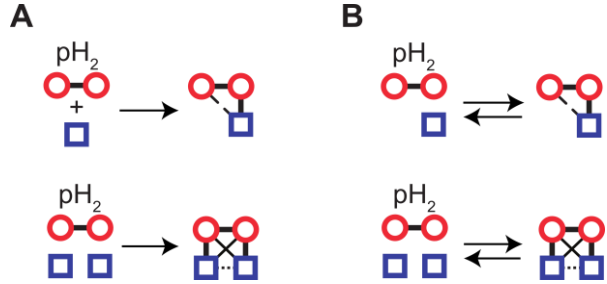

**Figure 1. Schematic view of hydrogenative (A, left) and non-hydrogenative (B, right) PHIP for a 3-spin-½ system with asymmetric couplings (top) and a 4-spin-½ system with symmetric couplings (bottom).** Here, we focus on hydrogenative PHIP in symmetric and asymmetric systems (A). The case of 4-spin-½ SABRE (B, bottom) was considered by Levitt in the seminal paper [14].

In the PASADENA case, upon the addition of pH$_2$ to an asymmetric molecular environment at high fields, $^1$H spins are weakly coupled. Since individual molecular hydrogenation events are distributed in time over the course of the hydrogenation reaction, the X and Y coherences (eq (1)) are lost, and the singlet spin state $\hat{\rho}_S^{A,B}$ is averaged to the so-called ZZ spin order [1]:

$$\hat{\rho}_{ZZ}^{A,B} = \frac{\hat{1}^N}{2^N} - \frac{1}{2^{N-2}} \hat{I}_Z^A \cdot \hat{I}_Z^B. \quad (3)$$

From now on, we will omit operator "hats" for simplicity.

It is common to transfer pH$_2$-derived spin alignment to proton and X-nuclear magnetization (e.g., $^{13}$C, $^{15}$N, $^{19}$F) for use as a MR imaging contrast agent [16–18], monitoring of chemical and enzymatic reactions [19,20], or for the purpose of analytical chemistry

[21]. Many of such spin-order transformations are represented by unitary transformations of the density matrix:

$$\rho(t) = U(t,t_0)\rho(t_0)U(t,t_0)^\dagger, \qquad (4)$$

where $\rho(t_0)$ is the density matrix at timt $t_0$, before the spin-order transfer (SOT), and $\rho(t)$ is the final density matrix, after the SOT. The unitary evolution operators $U(t,t_0)$, also known as propagators, can be found by solving the corresponding Liouville von-Neumann equation:

$$\frac{d}{dt}U(t,t_0) = -H(t)U(t,t_0) \qquad (5)$$

for a time-dependent Hamiltonian $H(t)$ and the initial condition $U(t_0,t_0) = 1^N$.

In this work, we discuss the transformation of the signlet state density matrix $\hat{\rho}_S^{A,B}$ and "PASADENA" density matrix $\hat{\rho}_{ZZ}^{A,B}$ to observable magnetization using general properties of unitary transformations [22,23] together with restrictions imposed by molecular symmetry [14].

## 2. Methods

### 2.1. Spin operators and observables

The general SOT from the initial spin state $\sigma_{\text{initial}}$ to the desired target spin state $\sigma_{\text{target}}$ under the action of propagator $U$ can be written as

$$U\sigma_{\text{initial}}U^\dagger = \sigma_{\text{final}} = \xi\sigma_{\text{target}} + \sigma_{\text{rest}} \qquad (6)$$

Where $\sigma_{\text{final}}$ is the final spin state, $\xi$ is the amplitude of the target spin state $\sigma_{\text{target}}$ and $\sigma_{\text{rest}}$ is the difference between $\sigma_{\text{final}}$ and $\xi\sigma_{\text{target}}$ that is not relevant for our considerations. We will use $\rho$ for density matrices and $\sigma$ for spin operators or traceless density matrices.

Since the propagator $U$ is unitary, the transformation (6) implies boundaries on the parameter $\xi$. There is no general way to determine all possible final states $\sigma_{\text{final}}$ for undefined $U$. However, it is possible to obtain boundary conditions for the amplitude $\xi \in [\xi_{\text{min}}, \xi_{\text{max}}]$ and a given $\sigma_{\text{initial}}$ and $\sigma_{\text{target}}$ in general.

We will define the spin operator of polarization of a single spin (e.g., A) as

$$\sigma_P^A = \frac{1}{2^{N-1}} I_Z^A \qquad (7)$$

and the spin operator of polarization of $N$ spins-½ as

$$\sigma_P^N = \frac{1}{2^{N-1}} \sum_{k=1}^{N} I_Z^k. \qquad (8)$$

Now it is straightforward to calculate a polarization, $P$, of one spin, or the average of many spins, using corresponding spin operators (7)(8):

$$P = \frac{\text{Tr}(\rho(t) \cdot \sigma_P)}{\text{Tr}(\sigma_P \cdot \sigma_P)}. \qquad (9)$$

Here, $\rho(t)$ is the density matrix of the system at the time of interest $t$.

In the same fashion, the amplitude $\xi$ of the state $\sigma_{\text{target}}$ for $\sigma_{\text{final}}$ after SOT can be evaluated as:

$$\xi = \frac{\text{Tr}(\sigma_{final} \cdot \sigma_{target})}{\text{Tr}(\sigma_{target} \cdot \sigma_{target})}. \qquad (10)$$

We will use $\xi$ in the following to report the maximum theoretically possible polariaztion ($\sigma_{target} = \sigma_P$).

### 2.2. No symmetry constraints [22]

The boundaries for the amplitude $\xi$ of the target state $\sigma_{\text{target}}$ after SOT (eq (6)) are

$$\xi_{\text{max}} = \|\sigma_{\text{target}}\|^{-1} \cdot \left(\Lambda_{\text{initial}}^\uparrow \cdot \Lambda_{\text{target}}^\uparrow\right), \qquad (11)$$

$$\xi_{\min} = \|\sigma_{\text{target}}\|^{-1} \cdot \left(\Lambda_{\text{initial}}^{\uparrow} \cdot \Lambda_{\text{target}}^{\downarrow}\right),$$

$$\|\sigma_{\text{target}}\| = \left(\Lambda_{\text{target}}^{\uparrow} \cdot \Lambda_{\text{target}}^{\uparrow}\right).$$

Here, $\Lambda_{\text{initial}}$ and $\Lambda_{\text{target}}$ are the eigenvalue vectors of the operators $\sigma_{\text{initial}}$ and $\sigma_{\text{target}}$. The arrows up ($\uparrow$) and down ($\downarrow$) indicate that these eigenvalues are sorted in an ascending or descending order. In general, $\xi_{\min} \leq \xi \leq \xi_{\max}$.

These boundaries arise because we assume all transformations to be unitary, and the initial and final states are given by Hermitian operators [22].

Because $\sigma_{\text{initial}}$ and $\sigma_{\text{target}}$ are traceless operators, the boundary parameters often have the same absolute value: $\xi_{max} = |\xi_{min}|$ unless otherwise noted.

2.3. *Symmetry constraints (SC)* [14,22,23]

When a system has a spin symmetry (i.e., groups of equivalent spins), only the states belonging to the same irreducible representations ($\Gamma$) of this group of symmetry $G$ can be mixed by unitary transformations [14,22,23]. In this case, the boundary conditions can be found as:

$$\xi_{\max}^{\text{SC}} = \|\sigma_{\text{target}}\|^{-1} \sum_{\Gamma} \left(\Lambda_{\text{initial}}^{\uparrow,\Gamma} \cdot \Lambda_{\text{target}}^{\uparrow,\Gamma}\right),$$

$$\xi_{\min}^{\text{SC}} = \|\sigma_{\text{target}}\|^{-1} \sum_{\Gamma} \left(\Lambda_{\text{initial}}^{\uparrow,\Gamma} \cdot \Lambda_{\text{target}}^{\downarrow,\Gamma}\right).$$
(12)

Where $\Lambda_X^{\uparrow,\Gamma}$ are vectors of eigenvalues of the operator $\sigma_X$ (X = initial or target) that correspond to eigenvectors belonging to the same $\Gamma$ and sorted in an ascending ($\uparrow$) or descending ($\downarrow$) order.

The transformation amplitude $\xi$ is bounded as $\xi_{\min} \leq \xi_{\min}^{\text{SC}} \leq \xi \leq \xi_{\max}^{\text{SC}} \leq \xi_{\max}$.

2.4. *Eigenvalues in the case of SC*

It is not trivial to define $\Lambda_X^{\uparrow,\Gamma}$ if SCs are present. To find the transformation of a density matrix $\sigma$ into a *group-symmetrized* basis, one needs to construct a symmetry group specific matrix $Q$ from orthonormal basis vectors $\vec{v}$. Each vector $\vec{v}$ must belong only to one irreducible representations $\Gamma$. Vectors $\vec{v}$ are written vertically. Let us enumerate these vectors in such a way that all vectors from the same $\Gamma$ stand next to each other: $Q = \left(\vec{v}_1^{\Gamma 1}, \vec{v}_2^{\Gamma 1}, \vec{v}_3^{\Gamma 1}, \ldots \vec{v}_1^{\Gamma 2}, \ldots \vec{v}_m^{\Gamma k}\right)$. Then, the matrix of spin state $\sigma$ ($\sigma_{\text{initial}}$ or $\sigma_{\text{target}}$) in the new basis $\sigma^Q$ can be found as

$$\sigma^Q = Q^{-1}\sigma Q.$$
(13)

There are three different situations for $\sigma^Q$:

1. **$\sigma^Q$ is diagonal.** When the predefined basis of the group $G$ coincides with the eigenstates of the operator $\sigma$, then $\sigma^Q$ is diagonal with eigenvalues on the diagonal. All coherences (off-diagonal elements) are zero. To find $\Lambda^{\uparrow or \downarrow,\Gamma}$ one has only to sort and enumerate the eigenvalues inside each $\Gamma$:

$$\sigma^Q = Q^{-1}\sigma Q = \begin{pmatrix} \Lambda_1^{\Gamma 1} & 0 & \cdots & 0 \\ 0 & \Lambda_2^{\Gamma 1} & & 0 \\ \vdots & & \ddots & \vdots \\ 0 & 0 & \cdots & \Lambda_m^{\Gamma k} \end{pmatrix}.$$
(14)

Here, $\Lambda_m^{\Gamma n}$ is an eigenvalue of $\sigma$ and $\vec{v}_m^{\Gamma n}$ its corresponding eigenvector belonging to irreducible representations $\Gamma n$.

2. **$\sigma^Q$ is $\Gamma$-block diagonal.** When $\sigma^Q$ has coherences only inside the same irreducible representations $\Gamma$, then $\sigma^Q$ is $\Gamma$-block diagonal matrix

$$\sigma^Q = Q^{-1}\sigma Q = \begin{pmatrix} \Lambda^{\Gamma 1} & 0 & \cdots & 0 \\ 0 & \Lambda^{\Gamma 2} & \cdots & 0 \\ \vdots & & \ddots & \vdots \\ 0 & 0 & \cdots & \Lambda^{\Gamma k} \end{pmatrix}. \tag{15}$$

Here, $\Lambda^{\Gamma m}$ are the corresponding blocks of irreducible representations $\Gamma m$. Because all vectors from one $\Gamma$ have the same symmetry, their superposition also has the same symmetry. It means that each block $\Lambda^{\Gamma m}$ should be additionally diagonalized, and resulting diagonal elements are corresponding eigenvalues $\Lambda_m^{\Gamma n}$.

3. **$\sigma^Q$ is not block diagonal.** The most general case is when there are off-diagonal elements between different irreducible representations.

$$\sigma^Q = Q^{-1}\sigma Q = \begin{pmatrix} \Lambda^{\Gamma 1} & C_{\Gamma 1}^{\Gamma 2} & \cdots & C_{\Gamma 1}^{\Gamma k} \\ C_{\Gamma 2}^{\Gamma 1} & \Lambda^{\Gamma 2} & \cdots & C_{\Gamma 2}^{\Gamma k} \\ \vdots & & \ddots & \vdots \\ C_{\Gamma k}^{\Gamma 1} & C_{\Gamma k}^{\Gamma 2} & \cdots & \Lambda^{\Gamma k} \end{pmatrix}. \tag{16}$$

In this situation, we will assume that the coherences $C_{\Gamma n}^{\Gamma m}$ are averaged to zero during the hydrogenation reaction due to magnetic field inhomogeneity and the different evolution time of each hydrogenated molecule. When such off-diagonal elements are removed ($C_{\Gamma n}^{\Gamma m} \coloneqq 0$), the $\sigma^Q$ is "$\Gamma$-block diagonal" and equivalent to eq (15). Hence, the consequent diagonalization and analysis is equivalent and described in "case 2".

In the following discussion, we use these three methods to find eigenvalues to evaluate $\xi$. A script is available in SI to evaluate $\xi$ for a different number of spins, symmetry, initial and target spin states (Matlab).

Below we will discuss some specific cases and demonstrate the effect of symmetry on PHIP and spin order transfer.

### 2.5. *Spin systems notations*

We use a notation that is slightly different from Pople's spin-system notation. The main idea is to distinguish the symmetries of the spin system. In addition, we fix X-spin to the target $^{13}$C nucleus. Let us consider some examples.

First, for us, "ABC" stands for a system with three chemically nonequivalent spins and only weak coupling is considered between spins. Second, according to Pople's notation, $^{12}C_2$-ethylene consists of four chemically and magnetically equivalent $^1$H spins, hence the spin symmetry is $A_4$. However, it does not reflect the permutation group symmetry of ethylene. Hence, we refer to spin symmetry of $^{12}C_2$-ethylene as AA'(AA').

## 3. Results

Note that the polariztions reported in the following are the upper limit of what can be transferred theoretically (as implied by the transformation mathematics).

### 3.1. *Parahydrogen spin-order transfer in a two spin-½ system*

#### 3.1.1. The symmetry of AB and $A_2$ systems

The simplest PHIP system consists of two spin-½ nuclei. If the protons of pH$_2$ after hydrogenation are magnetically and chemically nonequivalent (AB-system) – no symmetry constrains. The spins can be treated separately and the appropriate basis would be the *Zeeman* basis:

$$S^{AB} = \{|\alpha\alpha\rangle, |\alpha\beta\rangle, |\beta\alpha\rangle, |\beta\beta\rangle\}. \tag{17}$$

When the protons are magnetically equivalent, the system is called A₂ that imposis restrictions on the choise of the basis. Here singlet-triplet (S-T) states should be used:

$$S^{A_2} = \{|S\rangle, |T_+\rangle, |T_0\rangle, |T_-\rangle\}. \tag{18}$$

Among these two systems, only A₂ has nontrivial symmetry, which is C₂. In **Appendix A** we describe all relevant groups of symmetry. The transformation elements of C₂ group are identity transformation $E$ or null permutation "( )" and permutation of two protons $\binom{12}{21} = (21)$:

$$G^{12} = \{(\ ), (21)\}. \tag{19}$$

The C₂ group (or $G^{12}$) has two irreducible representations: even (gerade – "g") and odd (ungerade - "u"). The singlet state is the only member of odd irreducible representation $\Gamma^u$=B, while three triplet states are the members of $\Gamma^g$=A. In terms of sets, it can be written as

$$\begin{aligned} S_A^{12} &= \{|T_+\rangle, |T_0\rangle, |T_-\rangle\}, \\ S_B^{12} &= \{|S\rangle\}. \end{aligned} \tag{20}$$

Tables of characters and decomposition of spin states into irreducible representations are given for A₂, A₃, AA' (AA') systems in **Appendix A**.

### 3.1.2. pH₂ to magnetization in AB systems

Let us consider the transformation of $\sigma_{ZZ}^{A,B}$ spin order in an AB system (no symmetry constraints, eq (11)) to magnetization (eq (7)(8)):

$$\sigma_{ZZ}^{A,B} = -I_Z^A \cdot I_Z^B \rightarrow \begin{cases} \frac{1}{2} I_Z^{A,\text{or } B}, & |\xi| = 1, \\ \frac{1}{2}[I_Z^A + I_Z^B], & |\xi| = \frac{1}{2}, \\ \frac{1}{2}[I_Z^A - I_Z^B], & |\xi| = \frac{1}{2}. \end{cases} \tag{21}$$

This means that PASADENA spin order ($\sigma_{ZZ}^{A,B}$) can be transferred to 100% polarization of one spin, or 50% polarization of each spin. In the latter case, the net magnetization can be 50% per spin or zero (21).

The examples of spin-order transfer (SOT) sequences for direct polarization transfer to one spin are Selective Excitation of Polarization using PASADENA (SEPP)[24] and adiabatic-passage spin order conversion (APSOC)[25–27]. SOT for polarization transfer to two spins include out of phase echo (OPE)[28], only parahydrogen spectroscopy (OPSY)[28,29] and APSOC[25–27].

### 3.1.3. pH₂ to magnetization in A₂ systems

The situation is different for two magnetically equivalent spins A¹ and A² in an A₂ spin system. Symmetry constraints do not allow spin order conversion of $\sigma_S^{A,B}$ into net magnetization:

$$\sigma_S^{A^1,A^2} = -(\mathbf{I}^{A^1} \cdot \mathbf{I}^{A^2}) \rightarrow \frac{1}{2}I_Z^{A^1} + \frac{1}{2}I_Z^{A^2}, \quad |\xi^{SC}| = 0 \tag{22}$$

The only way to transfer polarization is to break the symmetry (A₂ → AB) that is happening e.g. during ALTADENA.

### 3.1.4. Limitation of the method: ALTADENA example

One of the first experiments that demonstrated spin order conversion of pH₂ was ALTADENA [7], using adiabatic magnetic field variation (AMFV). AMFV-induced spin order transfer in an AB, two spin-½ system results in the following transformation of $\sigma_S^{A,B}$ [1]:

$$\sigma_S^{A,B} = -(\mathbf{I}^A \cdot \mathbf{I}^B) \xrightarrow{\text{AMFV}} -I_Z^A \cdot I_Z^B \pm \frac{1}{2}(I_Z^A - I_Z^B). \tag{23}$$

The sign ($\pm$) depends on the relative chemical shift difference and the sign of $J$-coupling constant of the spins [1]. It follows that in ALTADENA, both spins acquire maximum polarization of 1 (see eq (7)), but the total (net) polarization of the molecule is zero.

The Hamiltonian of an AB system $H_{AB}$ is, in general, asymmetric (lacking permutation symmetry). However, at zero field ($B_0 = 0$), it has the same permutation symmetry as the Hamiltonian of $A_2$ system $H_{A_2}$:

$$H_{AB}(B_0 \neq 0) = -\nu_A^0 I_Z^A - \nu_B^0 \hat{I}_Z^B + J_{AB}(\mathbf{I}^A \cdot \mathbf{I}^B),$$

$$H_{AB}(B_0 = 0) = J_{AB}(\mathbf{I}^A \cdot \mathbf{I}^B), \tag{24}$$

$$H_{A_2}(B_0 \neq 0) = -\nu_A^0 \left(I_Z^{A^1} + I_Z^{A^2}\right) + J_{A^1 A^2}\left(\mathbf{I}^{A^1} \cdot \mathbf{I}^{A^2}\right).$$

This means that the initial and final symmetries of the Hamiltonian (and the spin system) are different and eq (12) cannot be used for the estimation of $\xi^{SC}$ (i.e., the symmetry changes during the experiment). We do not introduce a calculation method for such situations.

However, notice that the symmetry of the Hamiltonian (24) changes by introducing a magnetic field, while molecular symmetry does not change. This means that molecular symmetry does not have to coincide with the spin symmetry (or the symmetry of the nuclear spin Hamiltonian). The latter is essential for SOT, but molecular symmetry is essential for spin isomers (discussed below). Thus, there are at least three relevant symmetries: the Hamiltonian, interactions, and the spatial configuration of the molecule.

### 3.1.5. pH₂ on the a surface of a solid

It was predicted that the spin order of pH₂ after chemisorption, i.e. interaction with a surface, could be transformed into net magnetization even when the two spins have the same chemical shifts [30,31]. Note that there are a minimum of two requirements for PHIP via chemisorption: (a) the pH₂ nascent protons have to be chemically nonequivalent, and (b) if they split, there is a non-zero chance to reunite again with preserved quantum coherences. The main reason for spin order conversion is the difference in chemical shifts and intramolecular dipole-dipole (DD) interaction which is relevant on the surface. The Hamiltonian of such AB system at zero-field is

$$H_{AB}^{DD}(B_0 = 0) = J_{AB}(\mathbf{I}^A \cdot \mathbf{I}^B) + d(\theta, \varphi)\left(3I_Z^A \cdot I_Z^B - (\mathbf{I}^A \cdot \mathbf{I}^B)\right). \tag{25}$$

As a result, the state of the dihydrogens after chemisorption of pH₂ is expected to be a superposition of $\sigma_S^{A,B}$ and $\sigma_{ZZ}^{A,B}$ [31]:

$$\sigma_{S-DD}^{A,B} = -(P_{ZZ} - P_S)I_Z^A \cdot I_Z^B - P_S(\mathbf{I}^A \cdot \mathbf{I}^B). \tag{26}$$

Where P are the relative weights of the states $\sigma_S$ and $\sigma_{ZZ}$. This result was predicted for two AB spins with the Hamiltonian $H_{AB}^{DD}$ (eq (26)) by averaging the pH₂ derived initial spin state, $\sigma_S^{A,B}$, over hydrogenation time period.

We showed before that it is impossible to transfer $\sigma_S^{A,B}$ to total net magnetization of two spins in an $A_2$ system (eq (22)). However, it is possible to transfer $\sigma_{ZZ}^{A,B}$:

$$\sigma_{ZZ}^{A,B} = -I_Z^A \cdot I_Z^B \to \frac{1}{2}[I_Z^A + I_Z^B], \quad |\xi^{SC}| = 0.5, \tag{27}$$

even with symmetry constraints. Note that $|\xi^{SC}| = |\xi| = 0.5$ (compare eq (21) and eq (26)).

However, one should bear in mind that the mere feasibility of such transfer computed using the presented approach does not take into account whether or not there are interactions in the system that can be used for observation of the resulting spin order. Using solid echo sequences, SOT from $\sigma_{ZZ}^{A,B}$ to $\sigma_{P}^{A,B}$ was predicted for chemisorbed pH$_2$ [32,33].

### 3.2. Tranfer of pH$_2$ spin order to $^1$H magnetization in multispin systems

#### 3.2.1. No symmetry constraints

Now we will consider the transfer of either $\sigma_{ZZ}^{A,B}$ or $\sigma_{S}^{A,B}$ spin order into total spin magnetization, $\sigma_{P}^{N}$ (eq (8)), in N spin-½ systems with various topologies (**Figure 2**).

The highest level of polarization (i.e., the average polarization across all coupled spins) is possible if the system is in a pure singlet state $\sigma_{S}^{A,B}$ rather than in $\sigma_{ZZ}^{A,B}$ (**Figure 3 and 4** and **Appendix B, tables B1, B2**).

This situation is achieved when the S-T states are a (stationary) eigenbasis, i.e. when the J-couplings dominate the interactions. This can be achieved by adding pH$_2$ at low fields like in ALTADENA or via strong RF-pulses suppressing chemical shift evolution at high field [34]. Spin order transfer at low field can be realized using SLIC pulses and was demonstrated for molecular systems with up to five nonequivalent spins [35].

Interestingly, the maximum achievable polarization per molecule increases up to 400% (i.e. the equivalent of 4 spins polarized to 100%) if the system has 9 spins or more and the initial desity matrix is $\sigma_{S}^{A,B}$.

#### 3.2.2. With symmetry constraints

When a spin system has any permutation symmetries, the theoretically achievable proton polarization drops significantly. For example, in the A$_3$B$_2$C system of ethanol, the maximum average polarization is only 14.2% if pH$_2$ were added pairwise in positions A and B. If the ssymmetry contraints are relaxed so that six spins nonequivalent result, the maxium achievable polarization increases to 31.2% for $\sigma_{\text{initial}} = \sigma_{ZZ}^{A,B}$ and 50% for $\sigma_{\text{initial}} = \sigma_{S}^{A,B}$, respectively (**Figure 3**).

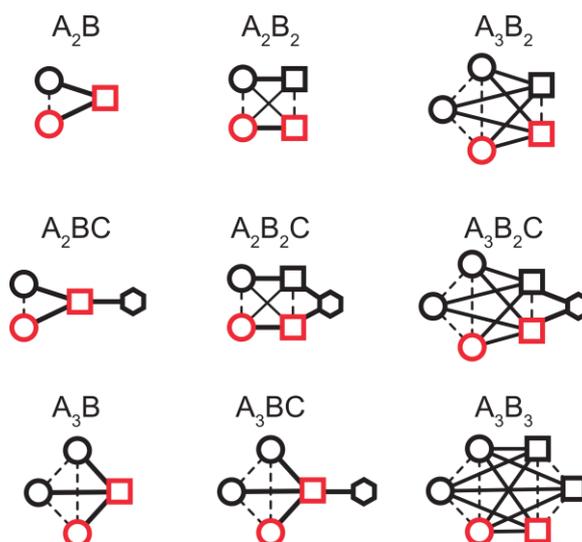

**Figure 2. Spin topologies considered for simulating the effect of symmetry on the transformation of pH2-derrived spin order (red) into observable polarization.** Red symbols indicate the pH$_2$-nascent spins, different lines indicate J-coupling constants, circles, squares, hexagons are spins of the same type.

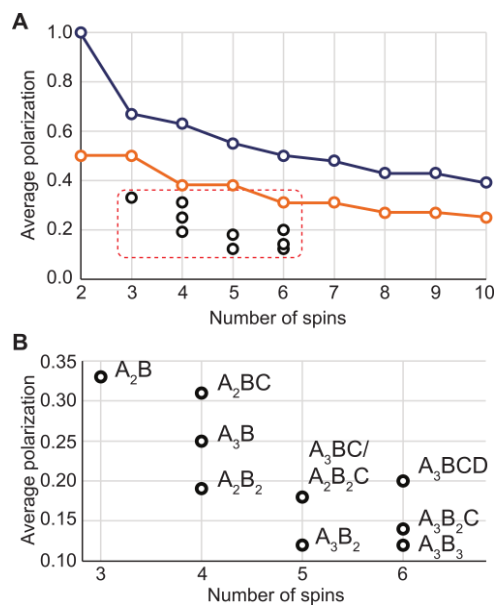

**Figure 3. Average polarization per spin that can be achieved theoretically by adding pH$_2$ to a molecule with 2 – 10 spins with (black) and without (orange, blue) symmetry contraints.** In general, higher polarization can be achieved if there are no constrains (compare black with orange and blue) and if the initial density matrix is $\sigma_S^{A,B}$ (blue) rather than $\sigma_{ZZ}^{A,B}$ (orange, compare blue and orange in A). We assumed pH$_2$ to be added in positions A and B. The reported values are given in **tables B1** and **B2 (Appendix B)**.

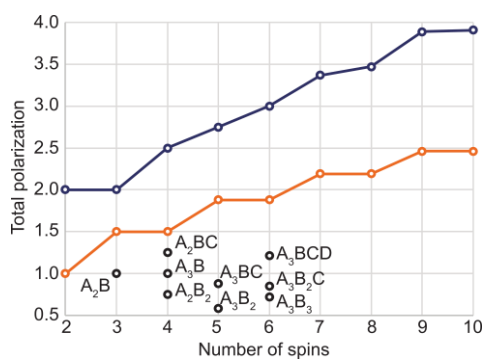

**Figure 4.** Sum of the average polarization – in units of one-spin polarization – that can be achieved theoretically by adding pH$_2$ to a molecule with 2 – 10 spins with (black, $\sigma_S^{A,B} \to \sigma_P^N$) and without symmetry constraints for $\sigma_S^{A,B} \to \sigma_P^N$ (blue) and $\sigma_{ZZ}^{A,B} \to \sigma_P^N$ (orange). If the polarization of all spins are summed up, up to ~ 400 % and above for more spins was obtained for large spin systems (blue). The reported values can be obtained from the average values given in **tables B1, B2 (Appendix B)**.

### 3.2.1. Nuclear spin isomers of H$_2$ and ethylene

**Dihydrogen**. We introduce nuclear spin isomers of molecules (NSIM), starting with H$_2$. The molecular symmetry group of H$_2$ is $D_{\infty h}$, while the permutation symmetry group of two spins is only C$_2$. The $D_{\infty h}$ symmetry include an infinite number of symmetry elements and is a product of C$_2$ and $C_\infty$ rotation groups, $S_\infty$ rotation-reflection and $C_V$ groups. For the sake of simplicity, and to exemplify NSIM, let us consider the C$_2$ symmetry only. The four nuclear spin states (sp) of H$_2$ can be grouped in two sets A and B (eq (20) and **Appendix A**): $A^{sp}$ (3 states, oH$_2$) and $B^{sp}$ (1 state, pH$_2$).

**Ethylene**. Ethylene is another example of a molecule with different NSIMs. Unlike H$_2$, however, the molecular symmetry group of ethylene is D$_{2h}$. Although the permutation symmetry group of spins is D$_2$ [36], it is helpful to use molecular symmetry to have a connection to corresponding rotational symmetries.

The permutation D$_2$ subgroup consists of 1 trivial and three nontrivial permutations that correspond to three orthogonal 180° rotations (**Figure 5**). In the D$_2$ symmetry group (see **Appendix A**), the 16 spin (sp) states of ethylene are grouped in 4 sets that correspond to A$^{sp}$, B$_1^{sp}$, B$_2^{sp}$ and B$_3^{sp}$ symmetries; seven states for A-symmetry set and 3 states for each of B-symmetry sets.

D$_{2h}$ symmetry group includes additional inversion operation (**i**) and its combinations with the above-mentioned 180° rotations. In this case, the decomposition of 16 spin states also results in 4 groups with additional symmetry indices: A$_g^{sp}$ (7 states), B$_{1u}^{sp}$ (3 states), B$_{2u}^{sp}$ (3 states) and B$_{3g}^{sp}$ (3 states). Seven states of A$_g^{sp}$ symmetry include five states with total spin 2 (A$_{g,2}^{sp}$) and two states with total spin 0 (A$_{g,0_1}^{sp}$ and A$_{g,0_2}^{sp}$). Each B group corresponds to three spin states of the same symmetry with the total spin 1.

The parity of spin states is even, and so is the parity of these four groups of symmetry. The rotational wavefunctions of ethylene are all of g symmetry, which leaves only four rotational (rot) symmetries A$_g^{rot}$, B$_{1g}^{rot}$, B$_{2g}^{rot}$, B$_{3g}^{rot}$.

The total (rotational and nuclear spin) wavefunction should have A-symmetry (either A$_g$ or A$_u$), therefore there are again four allowed combinations for ethylene: A$_g^{rot}$A$_g^{sp}$, B$_{1g}^{rot}$B$_{1u}^{sp}$, B$_{2g}^{rot}$B$_{2u}^{sp}$, B$_{3g}^{rot}$B$_{3g}^{sp}$.

We discuss the ethylene case in detail because it gives very good insights for the problem of polarization transfer between states of different symmetries.

Upon NSIM interconversion, the associated energy changes are by far larger than what is normally induced in NMR by RF pulses. A NSIM interconversion inevitably changes both rotational and spin states of ethylene. This is in strong contrast to non-symmetric molecules where the energy of spin flips is comparable to the strength of nuclear spin interactions. The large gap between H$_2$ rotational energies helps to enrich pH$_2$ state at low temperatures and is responsible for its long lifetime [3,5].

It may happen that some of the states belonging to different NSIM of polyatomic molecules have similar energies (in contrast to H$_2$ for which this is impossible). Such gateways are the basis of the quantum relaxation theory of NSIM interconversion [37]. For instance [38], for the B$_{1g}^{rot}$(J=23)B$_{1u}^{sp}$ and B$_{2g}^{rot}$(J=21)B$_{1g}^{sp}$ states of ethylene, the energy gap is "only" 46 MHz, i.e., within the reach of dipolar couplings. However, the energy of these states is 900 cm$^{-1}$ (1300 K) above the ground state of ethylene; therefore, their thermal population is low and the NSIM interconversion is also relatively slow despite efficient mixing of these states by the dipole-dipole interaction.

Unlike in H$_2$ (with C$_2$ symmetry), in ethylene, it seems possible to transfer singlet state and ZZ spin order into magnetization:

$$\sigma_S^{Am,An} = -\frac{1}{4}\mathbf{I}^{Am} \cdot \mathbf{I}^{An} \to \frac{1}{8}(I_Z^{A1} + I_Z^{A2} + I_Z^{A3} + I_Z^{A4})$$
$$\sigma_{ZZ}^{Am,An} = -\frac{1}{4}I_Z^{Am} \cdot I_Z^{An} \to \frac{1}{8}(I_Z^{A1} + I_Z^{A2} + I_Z^{A3} + I_Z^{A4})$$
(28)

with Am and An being one of four protons. Using the ethylene J-coupling constants to set the basis of spin states (**Appendix A** and ref [19] , $J_g$=1.07 Hz, $J_c$=11.47 Hz and $J_t$=17.78 Hz [39]), the following values were obtained for different hydrogenation sites: $|\xi_{S,cis}^{SC}| = |\xi_{S,trans}^{SC}| = |\xi_{S,gem}^{SC}| = 0.2$ and $|\xi_{ZZ,cis}^{SC}| = |\xi_{ZZ,trans}^{SC}| = |\xi_{ZZ,gem}^{SC}| = 0.225$. According to **Figure 5,** here cis corresponds to protons 1 and 4 (or 2 and 3), trans to 1 and 3 (or 2 and 4), gem to 1 and 2 (or 3 and 4). Note that using other simulations or isotop-labelling other (but similar) values were obtained and can be used [40]: $J_g$=2.23 – 2.39 Hz, $J_c$=11.62 – 11.66 Hz and $J_t$=18.99 – 19.03 Hz; [41]: $J_g$=2.5 Hz, $J_c$=11.6 Hz and $J_t$=19.1 Hz. However, because the constants of the same orde we did not compare the spin order transfer for different J-coupling values.

It may come as a surprise that efficent transfer of singlet spin order to polarization is feasible in a highly symmetric system like ethylene. So far, this transfer was demonstrated only in aligned media (nematic liquid crystals), where dipole-dipole spin-spin interactions remain, and the Hamiltonian of the system resembles $H_{AB}^{DD}$ (eq (25)). For such Hamiltonian as discussed above, instead of pure $\sigma_S^{Am,An}$, a mixture of $\sigma_S^{Am,An}$ and $\sigma_{ZZ}^{Am,An}$ should be considered and $\sigma_{ZZ}^{Am,An}$ spin order is observable in NMR.

Normally, the transition between two states with spin 0 that are represented by $A_{g,0_{1 \text{ or } 2}}^{sp}$ and $A_{g,2}^{sp}$ symmetries can not be observed by NMR. However, the theory applied here still allows the transfer of spin order to polarization. Let's have a closer look and find out why this is possible.

After hydrogenation with pH₂, six states with B-symmetry and two singlets of $A_g$ symmetry are populated under each hydrogenation scenario (**Table 1**). The average polarization that we considered in eq (28) depends on the populations of all symmetry states and is not straightforward to analyze. But let us instead consider the state $|2,2\rangle = |\alpha\alpha\alpha\alpha\rangle$ with the maximum for the system value of spin and spin projection of 2; one of five states of $A_g^{sp}$ symmetry. It means that if there is a way to transfer polarization from one of two $|0,0\rangle$ spin states ($A_{g,0_{1 \text{ or } 2}}^{sp}$) to the $|2,2\rangle$ state (both have the same $A_g^{sp}$ symmetry), ethylene hyperpolarization will be revealed. Now the question remains, how to achieve this, and if any existing spin order transfer methods, e.g. spin-lock induced crossing (SLIC)[42], adiabatic passage spin order conversion (APSOC)[27] or magnetic field cycling (MFC)[43] are suitable. This analysis goes beyond the scope of this paper and will be considered elsewhere.

**Table 1.** Relative populations of spin symmetries in ethylene after addition of pH₂ in germinal (gem), cis and trans position. Note that there are also coherences between $A_{g,0_1}^{sp}$ and $A_{g,0_2}^{sp}$ states, and between the two respectively populated B-symmetries (e.g. $B_{2u}^{sp}$ and $B_{3g}^{sp}$ in case of pos. = gem).

| Pos. | $A_{g,2}^{sp}$ | $A_{g,0_1}^{sp}$ | $A_{g,0_2}^{sp}$ | $B_{1u}^{sp}$ | $B_{2u}^{sp}$ | $B_{3g}^{sp}$ |
|---|---|---|---|---|---|---|
| gem | 0 | 0.2409 | 0.009 | 0 | 1/8 | 1/8 |
| cis | 0 | 0.1075 | 0.1425 | 1/8 | 0 | 1/8 |
| trans | 0 | 0.0266 | 0.2234 | 1/8 | 1/8 | 0 |

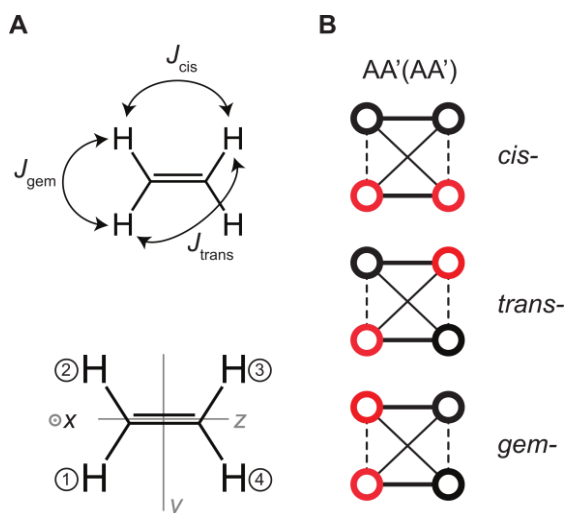

**Figure 5. Interactions and symmetry axis of ethylene .** A) Drawing of ethylene and its nuclear spin-spin couplings (*J*-couplings, top), numbering of the atomic positions and the cartheisan axis x, y, z. B) Graphs corresponding to the spin system AA'(AA') where pH₂ was added at cis-, trans- or geminal positions (red cicles). Different lines corresponds to different valus of spin-spin interactions.

### 3.3. PHIP-SAH and the transfer of pH₂ spin order to the magnetization of X-nuclei
#### 3.3.1. PHIP-SAH

Polarization transfer from pH₂ to X-nuclei also attracts significant attention in the context of hyperpolarized MRI applications. The lack of background signal and extended lifetime of polarization (compared to $^1$H) makes hyperpolarized MRI of X-nuclei highly interesting for biomedical applications, spearheaded by hyperpolariyed MRI of xenon and $^{13}$C-pyruvate. [16,44–47]

PHIP by sidearm hydrogenation (PHIP-SAH)[48,49] (**Figure 6**) attracted significant attention because it allowed polarization of acetate and pyruvate – the most commonly used contrast agents for hyperpolarized in vivo MRI. [50,51]

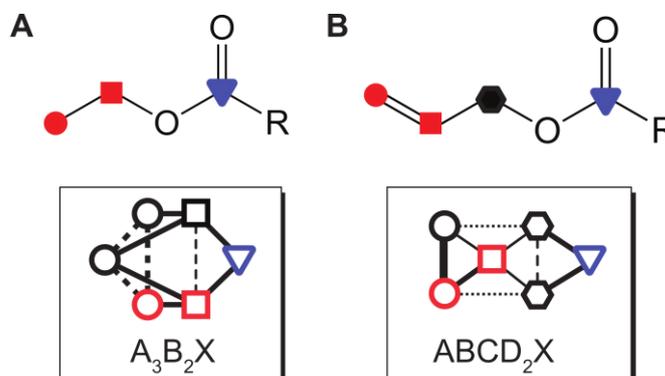

**Figure 6. Molecular structures (top) and spin topologies (bottom) of ethyl (A) and allyl (B) esters: of carboxylic acids - products of PHIP-SAH.** Different lines (bottom) represent different spin-spin interaction values.

### 3.3.2. No symmetry constraints

We considered the transfer of $\sigma_S^{A,B}$ (1) and $\sigma_{ZZ}^{A,B}$ (3) to X-nuclear polariaztion ($\sigma_P^X$) (7) without symmetry constraints (**Table 2**). In both cases, 100% polarization can be achieved: $|\xi| = 1$. This, for example, was predicted for hydrogenation of perdeuterated 1-$^{13}$C-vinyl-acetate-$d_6$ [9]. More than 50% polarization was achieved on $^{13}$C in the system consisting of three nonequivalent spin-½ and six spin-1 nuclei ($^2$H). The direct loss of polarization is due to S-T$_0$ mixing of pH$_2$–derived hydrogens at the catalyst and relaxation during hydrogenation [52].

### 3.3.3. Symmetry constraints

If some other (i.e., non nascent pH$_2$) protons possess any symmetry, still, 100% polarization transfer can be achieved on X (theoretically). This situation is realized e.g. in 1-$^{13}$C-allyl-pyruvate, an ABCD$_2$X system.

If one of the pH$_2$-nascent protons ends up in a symmetric site of the product, the maximum polarization that can be tranferd to X is reduced to 75% for A$_2$BX ($|\xi^{SC}| =3/4$) and 66.(6)% for A$_3$BX ($|\xi^{SC}| =2/3$). Again, the number of the "other" protons and their symmetry does not play a role.

If both pH$_2$ spins bind to two different symmetric spin sites, like A$_3$B$_2$X, A$_3$B$_2$C$_2$X, the maxium polarization that can be transferd to X is further reduced to 50% ($|\xi^{SC}| = 0.5$). This situation is found in ethyl- and propyl pyruvate e.g.

In the literature, about 20% $^{13}$C-polarization was reported on ethyl pyruvate. Here, pH$_2$ was added to vinyl pyruvate at high field and spin order was transferred to $^{13}$C using INEPT [53]. About 20-35% $^{13}$C-polarization was reported in a similar experiment, where pH$_2$ was added at low field and after magnetic field variation detection took place at high field [43,54]. However, one should remember that, as with the ALTADENA case discussed above, the Hamiltonian of the system changes during the magnetic field variation.

It is interesting to note that it is not possible to transfer spin order in an A$_3$X system if two of the A spins are in the singlet state($|\xi^{SC}| = 0$ for $\sigma_S^{A_i,A_j} \to \sigma_P^X$). If they are in the "reduced" singlet state $\sigma_{ZZ}$, however, trasnfer is possible($|\xi^{SC}| =1/3$ for $\sigma_{ZZ}^{A_i,A_j} \to \sigma_P^X$). For example, as we discussed before, pH$_2$ derived spin order after chemisorption is partially in ZZ state.

**Table 2.** Polarization transfer from pH$_2$ ($\sigma_S^{A,B}$) to an X nucleus ($\sigma_P^X$). One hydrogen of pH$_2$ is in A, the other is in B position.

| Type of the system | $\xi_{max}$ from $\sigma_S^{A,B} = \xi\sigma_P^X + \sigma_{rest}$ | Examples of molecule, R=acetate, pyruvate |
|---|---|---|
| ABX, ABCX, ABCDX, ABCDEX,… | 1 | 1-$^{13}$C-vinyl-R |
| ABCD$_2$X | | 1-$^{13}$C-allyl-R |
| AA'BB'X | 1 | 1-$^{13}$C-ethylene |
| A$_2$BX | ¾ | |
| A$_2$BC$_2$X | ¾ | |
| A$_3$BX | 2/3 | |
| A$_2$B$_2$X | 9/16 | |
| A$_3$B$_2$X | ½ | 1-$^{13}$C-ethyl-R |
| A$_3$B$_2$C$_2$X | ½ | 1-$^{13}$C-propyl-R |

### 3.3.4. Double hydrogenation

Now we turn to the question: is it beneficial to add two pH$_2$ molecuels to one target? For example, 1-$^{13}$C-ethynyl pyruvate is transformed into 1-$^{13}$C-ethyl pyruvate, an A$_3$B$_2$X system, by double hydrogenation (addition of two pH$_2$s). Likewise, 1-$^{13}$C-propargyl pyruvate becomes 1-$^{13}$C propyl pyruvate, an A$_3$B$_2$C$_2$X system, upon addition of two pH$_2$ (pH$_2$ is added to A and B in both cases).

Let us assume that the hydrogenation reaction is so fast that only the $|SS\rangle\langle SS|$ spin state is populated, which can be written as

$$\sigma_{S|S}^{A1,B1|A2,B2} = -\frac{1}{2^{N-2}}(\mathbf{I}^{A1} \cdot \mathbf{I}^{B1}) - \frac{1}{2^{N-2}}(\mathbf{I}^{A2} \cdot \mathbf{I}^{B2}) + \frac{1}{2^{N-4}}(\mathbf{I}^{A1} \cdot \mathbf{I}^{B1})(\mathbf{I}^{A2} \cdot \mathbf{I}^{B2}) \quad (29)$$

The theorerically maximum transfer from $\sigma_{S|S}^{A1,B1|A2,B2}$ to X-nuclear polarization in A$_3$B$_2$X and A$_3$B$_2$C$_2$X systems is $|\xi^{SC}| = 1/4$: 2 times lower than for the single hydrogenation. Note that it is also system specific, e.g. for A$_2$B$_2$X system double hydrogenation results in $|\xi^{SC}| = 3/4$, while $|\xi^{SC}| = 9/16$ is a maximum predicted for a single hydrogenation (**Tab. 2**).

In reality, however, there will be a finite time between the first and second hydrogeatnion, such that the system will start to evolve. As a result, the final state will be different than $|SS\rangle\langle SS|$·and the polarization of X nucleus will also be different.

The situation is similar for multiple hydrogenations in different positions as, for example, in trivinyl orthoacetate.[55] If we simplify the product to two ethyl groups, the system becomes (A$_3$B$_2$)(A$_3$B$_2$)'X, where one pH$_2$ is part of the A$_3$B$_2$ subsystem, and the other is part of (A$_3$B$_2$)' subsystem. In this case, polarization transfer amplitude $|\xi^{SC}| = 5/16$ was predicted, while for a single hydrogenation (yielding an A$_3$B$_2$X system), it was $|\xi^{SC}| = 1/2$.

### 3.4. Examples of isotopic and chemical symmetry breaking

#### 3.4.1. Ethylene

Ethylene produced by adding pH$_2$ to acetylene in isotropic environment does not demonstrate any enhanced observable magnetization. The spin state of pH$_2$ should be converted into ZZ-state instead: this was demonstrated after hydrogenation and subsequent dissolution of ethylene in a liquid crystal [19]. However, one can imaging to do it in different order. First, generate the ZZ-state of pH$_2$ derived H$_2$, that was estimated for solid catalyst [30,31]. Second, hydrogenate acetylene using this H$_2$.

In genera, the pH$_2$-derived protons are observed only when they are attached to chemically or magnetically inequivalent sites that for ethylene can be achieved at least in two ways:

- The two pairs of hydrogens in ethylene can be made magnetically nonequivalent by $^{13}$C labeling. In case of single-sided $^{13}$C labeling system symmetry drops down to C$_2$ and polarization transfer is possible to $^1$H or $^{13}$C nuclei. In addition, chemical shifts of two gem pairs of protons are different.
- The other way to break the ethylene symmetry is a chemical reaction. So, polarized ethylene gas bubbled through a CCl$_4$ solution of perfluoro(para-tolylsulfenyl) chloride (PTSC) yields an asymmetric PTSC/ethylene adduct [19]. As a result, a normal PASADENA spectrum can be obtained.

### 3.4.2. Fumarate and maleate

Fumarate and maleate are two metabolites with symmetry imposed spin order transfer restrictions; the solutions are also the same. The symmetry can be broken by a $^{13}$C labeling [56,57] or as a result of chemical reaction: "hyperpolarized" fumarate was converted by fumarase to asymmetric malate revealing itself in PASADENA spectrum [58].

Dimethyl ether of maleate (and fumarate, other popular PHIP molecule) has Cs symmetry. However, pH$_2$-derived protons are magnetically inequivalent because of interaction with two CH$_3$ groups and spin order of pH$_2$ can be accessed with RF pulses [56] or magnetic field variation [59].

## 4. Discussion

We considered several cases of polarization transfer from pH$_2$ to proton and X nuclei magnetization using the methods introduced in Refs. [14,22,23]. The approach used here helps to provide some general answers to several nonintuitive questions. However, a few situations remain unclear and may indicate some limitations of the presented theory. Namely,

**Q1.** How to estimate maximum polarization transfer from a state which is not diagonal in the basis of the system's symmetry?

**Q2.** How to estimate polarization transfer in systems that experience symmetry change during the polarization transfer, e.g. A$_2$→AB during magnetic field variation in ALTADENA experiment? Is there a general solution for an N-spin system?

**Discussion of Q1.** This situation corresponds to the third case of a $\sigma^Q$-diagonalization (eq (16)) as described in *methods*. For example, in an A$_2$BX system, the basis consists of the functions $|Mkl\rangle$ where $|M\rangle$ is one of S-T basis functions and $|k\rangle$, $|l\rangle$ are spin up and down, $|\alpha\rangle$ and $|\beta\rangle$. In this basis, $\sigma_S^{A1,B}$ is not diagonal. Instead, proejcting this state on the symmetry basis results in $\frac{1}{2}\left[\sigma_{ZZ}^{A1,B} + \sigma_{ZZ}^{A2,B}\right]$, meaning that we lose part of the initial spin order and potentially underestimate the level of polarization transfer.

**Discussion of Q2.** This problem was discussed in the context of ALTADENA, but it is also very important for magnetic field variation e.g. in PHIP-SAH. Let us consider a simple ABX system. At low fields, when proton chemical shift difference can be neglected, the ABX system becomes equivalent to an A$_2$X system meaning that for protons S-T basis is more appropriate at low fields. $\sigma_S^{A,B}$ is the initial state of the system after pH$_2$ addtion ($\sigma_{\text{initial}} = \sigma_S^{A,B}$). Then, we increase the field slowly so that the system changes from A$_2$X to ABX ans basis from S-T to Zeeman. This means that the symmetry basis of the system before and after (and during!) the transformation is different. The theory presented here can nob be applied.

Although in three-spin systems, we can still reach 100% X nuclear polarization, we again could underestimate the efficiency of polarization transfer in more complex systems.

It looks as if the methodology used here for the static high magnetic field can be translated to the low fields (and zero fields). However, the basis will be system symmetry specific and, in addition, will depend on the *J*-coupling network.

Assissing the validity of this approach is not straight forward. To date, however, experimental results have not contradicted the calcuatled results presented here.

## 5. Conclusion

The mathematical framework presented here allows to determine an upper limit for the polarization transfer from pH$_2$ to X-nuclei or other protons with an emphasis on the effect of molecular of spin symmetry. Solutions were presented for the most current experimental situations, although some more compex cases remain unaddressed. This method may serve as a first check to estimate if and how much polarization transfer is possible in a given situation. Naturally, identifying and optimizion a dedicated transfer strategy is the following essential step which ist not addressed here.

**Supplementary Materials:** All used Matlab scripts together with MOIN spin library [60] (.zip) are available online or on request.

**Author Contributions:** ANP conceptualization, software, DAB visualization, ANP, DAB original draft, ANP, DAB, IVK investigation, All review and editing, funding acquisition. All authors have read and agreed to the published version of the manuscript.

**Acknowledgments and funding:** We acknowledge funding from German Federal Ministry of Education and Research (BMBF) within the framework of the e:Med research and funding concept (01ZX1915C), DFG (PR 1868/3-1, HO-4602/2-2, HO-4602/3, GRK2154-2019, EXC2167, FOR5042, SFB1479, TRR287), Kiel University and the Faculty of Medicine. MOIN CC was founded by a grant from the European Regional Development Fund (ERDF) and the Zukunftsprogramm Wirtschaft of Schleswig-Holstein (Project no. 122-09-053). DAB acknowledges support from the Alexander von Humboldt Foundation in the framework of the Sofja Kovalevskaja Award. I.V.K. acknowledges the Russian Ministry of Science and Higher Education (grant no. 075-15-2020-779 and 075-15-2021-580) for financial support. We are grateful to Dmitry Budker for the discussion and editing of the manuscript.

**Conflicts of Interest:** The authors declare no conflict of interest.

## Appendix A

### 1. A$_2$ two spin-½ system, C$_2$ group

The number of states is $2^2=4$.

The basis for two equivalent spins can be divided into two groups with total spin $I^{\text{tot}}$ of 1 (three states) and 0 (one state) also known as singlet-triplet (S-T) basis. This can be derived formally by finding eigenfunctions and eigenvalues ($\lambda$) of cyclic permutation operator $\begin{pmatrix} 1 & 2 \\ 2 & 1 \end{pmatrix} = (21)$. **C**$_2$ permutation group of A$_2$ system consists of two permutations, {(),(21)}. In the matrix form written in the Zeeman basis ($|\alpha\alpha\rangle$, $|\alpha\beta\rangle$, $|\beta\alpha\rangle$, $|\beta\beta\rangle$):

$$() = E = \begin{pmatrix} 1 & 0 & 0 & 0 \\ 0 & 1 & 0 & 0 \\ 0 & 0 & 1 & 0 \\ 0 & 0 & 0 & 1 \end{pmatrix},$$
$$(21) = C_2 = \begin{pmatrix} 1 & 0 & 0 & 0 \\ 0 & 0 & 1 & 0 \\ 0 & 1 & 0 & 0 \\ 0 & 0 & 0 & 1 \end{pmatrix}.$$
(A1)

Eigenvalues of $C_2$ are given as a superscript to the corresponding wavefunctions:

$$\lambda = 1, \text{group } 1 \in A$$
$$|1,+1\rangle^1 = |T_+\rangle = |\alpha\alpha\rangle,$$
$$|1,0\rangle^1 = |T_0\rangle = \frac{|\alpha\beta\rangle+|\beta\alpha\rangle}{\sqrt{2}},$$
(A2)

$$|1,-1\rangle^1 = |T_-\rangle = |\beta\beta\rangle,$$

$$\lambda = -1, \text{ group } 2 \in B$$

$$|0,0\rangle^{-1} = |S_0\rangle = \frac{|\alpha\beta\rangle - |\beta\alpha\rangle}{\sqrt{2}}.$$

We indicate spin states by the total spin $I^{\text{tot}}$ and its projection $I_Z^{\text{tot}}$ as $|I^{\text{tot}}, I_Z^{\text{tot}}\rangle^\lambda$ and/or by using Zeeman basis, $|\alpha\rangle$ and $|\beta\rangle$.

**Table A1.** *Table of characters for A₂ two spin-½ system (C₂ group).*

|   | E | C₂ |
|---|---|---|
| **A** | 1 | 1 |
| **B** | 1 | -1 |
| **SpinRep = 3A + B** | 4 | 2 |

Therefore, there are three symmetric and one asymmetric states with respect to (12) permutation (or rotation about 180 degrees, C₂). It follows from both eq A1 and Table A1. Therefore, the basis for A₂ two spin-½ system consists of 2 sets (S) with multiplicity 3 and 1 (SpinRep = 3A + B):

$$S_A^{12} = \{|T_+\rangle, |T_0\rangle, |T_-\rangle\},$$
$$S_B^{12} = \{|S_0\rangle\}. \tag{A3}$$

To calculate characters for spin permutations (SpinRep) in Table A1, one can (i) write matrix of permutation and (ii) calculate trace. For an identity transformation, ()=E, for N spin-½ character is

$$\chi_{\text{SpinRep}}(E) = \text{Tr}(E) = 2^N. \tag{A4}$$

Analogously (eq A1) $\chi_{\text{SpinRep}}(C_2) = \text{Tr}(C_2) = 2$.

For any character $X$ of a representation $T = \oplus_i T_i$ which is superposition of irreducible representations of the same group, the multiplicity $n_k$ of the irreducible representation $T_i$ is given by

$$n_k = \frac{1}{\Omega_G} \sum_g X(g)^* \chi_k(g). \tag{A5}$$

Here "*" is a complex conjugate and $\Omega_G = \sum_g \chi_k(g)^* \chi_k(g)$. Eq A5 is useful to decompose the SpinRep line into a sum of characters for irreducible representations. So, for A₂ system SpinRep = 3A + B.

### 2. A₃ three spin-½ system, C₃ group

The number of states is $2^3 = 8$.

The basis of three equivalent spins can be grouped on three groups with total spin $I^{\text{tot}}$ of 3/2 (4 states), 1/2 (2 states) and 1/2 (2 states). Here also to distinguish groups, we introduce eigenvalues for cycling permutation operator $\begin{pmatrix}123\\231\end{pmatrix} = (231)$ and its values are given as a superscript to the corresponding wavefunctions.

$$I^{\text{tot}} = 3/2, \lambda = 1, \text{ group } 1 \in A$$

$$\left|\tfrac{3}{2}, +\tfrac{3}{2}\right\rangle^1 = |\alpha\alpha\alpha\rangle,$$

$$\left|\tfrac{3}{2}, +\tfrac{1}{2}\right\rangle^1 = \tfrac{1}{\sqrt{3}}(|\alpha\alpha\beta\rangle + |\alpha\beta\alpha\rangle + |\beta\alpha\alpha\rangle),$$

$$\left|\tfrac{3}{2}, -\tfrac{1}{2}\right\rangle^1 = \tfrac{1}{\sqrt{3}}(|\alpha\beta\beta\rangle + |\beta\alpha\beta\rangle + |\beta\beta\alpha\rangle)$$

$$\left|\tfrac{3}{2}, -\tfrac{3}{2}\right\rangle^1 = |\beta\beta\beta\rangle, \tag{A6}$$

$$I^{\text{tot}} = 1/2, \lambda = e^{\frac{i2\pi}{3}} = e^{+i\theta}, \text{ group } 2 \in E_1$$

$$\left|\tfrac{1}{2}, +\tfrac{1}{2}\right\rangle^{e^{+i\theta}} = \frac{|\alpha\alpha\beta\rangle + e^{-i\theta}|\alpha\beta\alpha\rangle + e^{+i\theta}|\beta\alpha\alpha\rangle}{\sqrt{3}},$$

$$\left|\tfrac{1}{2}, -\tfrac{1}{2}\right\rangle^{e^{+i\theta}} = \frac{|\beta\beta\alpha\rangle + e^{-i\theta}|\beta\alpha\beta\rangle + e^{+i\theta}|\alpha\beta\beta\rangle}{\sqrt{3}},$$

$$I^{\text{tot}} = 1/2, \lambda = e^{-\frac{i2\pi}{3}} = e^{-i\theta}, \text{group } 3 \in E_2$$

$$\left|\tfrac{1}{2}, +\tfrac{1}{2}\right\rangle^{e^{-i\theta}} = \frac{|\alpha\alpha\beta\rangle + e^{+i\theta}|\alpha\beta\alpha\rangle + e^{-i\theta}|\beta\alpha\alpha\rangle}{\sqrt{3}},$$

$$\left|\tfrac{1}{2}, -\tfrac{1}{2}\right\rangle^{e^{-i\theta}} = \frac{|\beta\beta\alpha\rangle + e^{+i\theta}|\beta\alpha\beta\rangle + e^{-i\theta}|\alpha\beta\beta\rangle}{\sqrt{3}}.$$

The C₃ permutation group $G = \{\binom{123}{123}, \binom{123}{231}, \binom{123}{312}\} = \{(), (+\theta), (-\theta)\}$ of A₃ system consists of three permutations: the trivial identity permutation, permutation or "+θ" rotation and "–θ" rotation, with $\theta = \frac{2\pi}{3}$.

**Table A2.** Table of characters for A₃ three spin-½ system (C₃ group). Here $\theta = \frac{2\pi}{3}$. Note the difference between three different "E" here. The characters for spin representations (SpinRep) are filled using eq A4 and following discussions. Any permutation (or rotation) will leave only states $|\alpha\alpha\alpha\rangle$ and $|\beta\beta\beta\rangle$ on the diagonal. It means that the sum of diagonal elements and corresponding character value is 2.

|  | E | $C_3^1$ | $C_3^2$ |
|---|---|---|---|
| A | 1 | 1 | 1 |
| E₁ | 1 | $e^{i\theta}$ | $e^{-i\theta}$ |
| E₂ | 1 | $e^{-i\theta}$ | $e^{i\theta}$ |
| **SpinRep = 4A + 2E₁+2E₂** | 8 | 2 | 2 |

Summarizing, there are 4 symmetric (g) states and two pairs of rotationally symmetric states (E_i, $|S\rangle$) states. It follows from both: eq A6 and Table A2. Therefore, the basis for A₃ three spin-½ system consists of 3 sets with multiplicity 4, 2 and 2 (SpinRep = 4A + 2E₁+2E₂):

$$S_A^{123} = \left\{\left|\tfrac{3}{2}, +\tfrac{3}{2}\right\rangle^1, \left|\tfrac{3}{2}, +\tfrac{1}{2}\right\rangle^1, \left|\tfrac{3}{2}, -\tfrac{1}{2}\right\rangle^1, \left|\tfrac{3}{2}, -\tfrac{3}{2}\right\rangle^1\right\},$$

$$S_{E_1}^{123} = \left\{\left|\tfrac{1}{2}, +\tfrac{1}{2}\right\rangle^{e^{+i\theta}}, \left|\tfrac{1}{2}, -\tfrac{1}{2}\right\rangle^{e^{+i\theta}}\right\}, \quad (A7)$$

$$S_{E_2}^{123} = \left\{\left|\tfrac{1}{2}, +\tfrac{1}{2}\right\rangle^{e^{-i\theta}}, \left|\tfrac{1}{2}, -\tfrac{1}{2}\right\rangle^{e^{-i\theta}}\right\}$$

### 3. A₄ four spin-½ system: C₄ group example

The number of states is $2^4 = 16$.

The basis of four spins can be grouped on 4 groups with total spin $I^{\text{tot}}$ of 2 (five states), 1 (3 groups, each consists of 3 states), and 0 (two groups, each one state).

C₄ permutation group $G = \{\binom{1234}{1234}, \binom{1234}{2341}, \binom{1234}{3412}, \binom{1234}{4123}\} = \{(), (2341), (3412), (4123)\}$ of A₄ system consists of four permutations: ( ) – trivial identity permutation and three cyclic permutations that are equivalent to rotation of a square by 90°, 180° and 270° around the center axis perpendicular to its plane. Note that only 180° rotation can be represented as two consequent permutations (3412) = (13)(24).

**Table A3.** Table of characters for A₄ four spin-½ system (C₄ group). The characters for SpinRep are filled using eq A4 and following discussion. Any rotations will leave states $|\alpha\alpha\alpha\alpha\rangle$ and $|\beta\beta\beta\beta\rangle$ on diagonal. All other states are changing after an odd number of cyclic permutations. Hence, character for $C_4$ and $(C_4)^3$ is only 2. 180° rotation ((13)(24) permutation) does not change also $|\alpha\beta\alpha\beta\rangle$ and $|\beta\alpha\beta\alpha\rangle$ states. Hence the corresponding character is 2+2=4. It means that the sum of diagonal elements and corresponding character values are 4 for each rotation (permutation).

|  | E | $C_4$ | $C_2 = (C_4)^2$ | $(C_4)^3$ |
|---|---|---|---|---|
| A | +1 | +1 | +1 | +1 |
| B | +1 | -1 | +1 | -1 |
| E$_1$ | +1 | +i | -1 | -i |
| E$_2$ | +1 | -i | -1 | +i |
| SpinRep = 6A + 4B+3E$_1$+3E$_2$ | 16 | 2 | 4 | 2 |

### 4. AA'(AA') four spin-½ system: D$_2$ group (spin symmetry of ethylene)

The number of states is $2^4$ = 16.

The basis of four spins can be grouped into 4 groups with total spin $I^{tot}$ of 2 (five states), 1 (3 groups, each consists of 3 states), and 0 (two groups, each one state).

D$_2$ permutation group $G$ = {$\begin{pmatrix}1234\\1234\end{pmatrix}$, $\begin{pmatrix}1234\\2143\end{pmatrix}$, $\begin{pmatrix}1234\\3412\end{pmatrix}$, $\begin{pmatrix}1234\\4321\end{pmatrix}$}= {(), (21)(43), (31)(42), (41)(32)} of AA'(AA') system consists of four permutations: ( ) – trivial identity permutation and three pair-wise permutations that are equivalent to rotations of the rectangle by 180° around three orthogonal axes, which are orthogonal to the plane of the rectangle and (or) its edges. We do not write here corresponding basis for general D$_2$ group which is equivalent to D$_{2h}$ discussed below.

**Table A4.** Table of characters for AA'(AA') four spin-1/2 system (D$_2$ group). The characters for SpinRep are filled using eq A4 and following discussion. Any rotations will leave states $|\alpha\alpha\alpha\alpha\rangle$ and $|\beta\beta\beta\beta\rangle$ on diagonal. In addition, states $|\alpha\alpha\beta\beta\rangle$ and $|\beta\beta\alpha\alpha\rangle$ do not change by the action of (21)(43) permutation. For two other rotations, one can also write the corresponding two states. It means that the sum of diagonal elements and corresponding character values are 4 for each rotation (permutation).

| | E | $C_2(z)$ | $C_2(y)$ | $C_2(x)$ |
|---|---|---|---|---|
| **A** | +1 | +1 | +1 | +1 |
| **B$_1$** | +1 | +1 | -1 | -1 |
| **B$_2$** | +1 | -1 | +1 | -1 |
| **B$_3$** | +1 | -1 | -1 | +1 |
| **SpinRep = 7A + 3B$_1$+3B$_2$+3B$_3$** | 16 | 4 | 4 | 4 |

### 5. AA' (AA') four spin-½ system, D$_{2h}$ group (molecular symmetry of ethylene)[19]

The number of states is $2^4$ = 16.

The basis of four spins can be grouped into 4 groups with total spin $I^{tot}$ of 2 (five states), 1 (3 groups, each consists of 3 states), and 0 (two groups, each one state).

D$_{2h}$ group is a direct product of D$_2$ and C$_i$ groups. D$_2$ part consists of 4 spin permutations $G(D_2)$ = {(), (12)(34), (13)(24), (14)(23)}. An addition of inversion operator of C$_i$ results in four additional transformations {$i, \sigma(xy), \sigma(xz), \sigma(yz)$}: inversion "i" and three mirror $\sigma$-planes: xy, xz or yz (**Figure A1 and 4**).

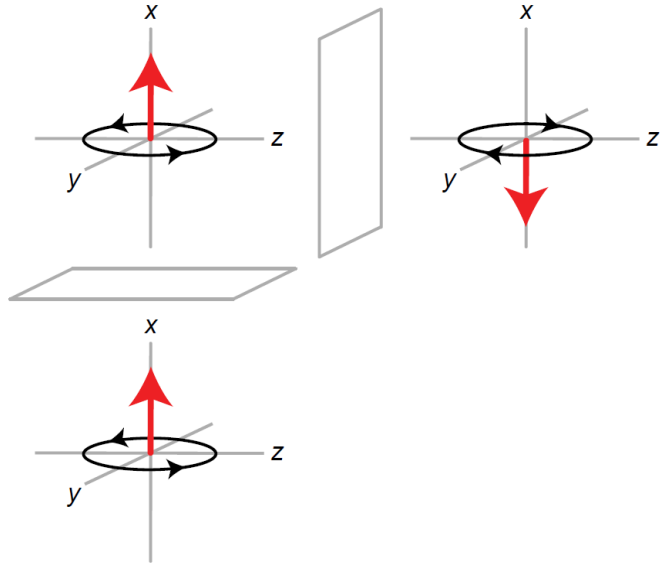

**Figure A1. Action of a mirror plane on an axial vector (spin or magnetic dipole).** When axial vector is perpendicular to the mirror plane, it does not change upon reflection. If, however, it is in oriented along the mirror plane, it changes its orientation upon reflection.[61]

The eigenvectors for ethylene were found before[19] and are given in eq (A10) without change.

$I^{tot} = 2$, group 1 $\in$ A$_g$

$|2,+2\rangle^g = |\alpha\alpha\alpha\alpha\rangle$,

$|2,+1\rangle^g = \frac{|\alpha\alpha\alpha\beta\rangle+|\alpha\alpha\beta\alpha\rangle+|\alpha\beta\alpha\alpha\rangle+|\beta\alpha\alpha\alpha\rangle}{2}$,

$|2,0\rangle^g = \frac{|\alpha\alpha\beta\beta\rangle+|\alpha\beta\alpha\beta\rangle+|\alpha\beta\beta\alpha\rangle+|\beta\alpha\beta\alpha\rangle+|\beta\beta\alpha\alpha\rangle+|\beta\alpha\alpha\beta\rangle}{\sqrt{6}}$,

$|2,-1\rangle^g = \frac{|\beta\beta\beta\alpha\rangle+|\beta\beta\alpha\beta\rangle+|\beta\alpha\beta\beta\rangle+|\alpha\beta\beta\beta\rangle}{2}$,

$|2,-2\rangle^g = |\beta\beta\beta\beta\rangle$,

$I^{tot} = 1$, group 2 $\in$ B$_{1u}$

$|1,+1\rangle^{1u} = \frac{-|\alpha\alpha\alpha\beta\rangle-|\alpha\alpha\beta\alpha\rangle+|\alpha\beta\alpha\alpha\rangle+|\beta\alpha\alpha\alpha\rangle}{2}$,

$|1,0\rangle^{1u} = \frac{|\alpha\alpha\beta\beta\rangle-|\beta\beta\alpha\alpha\rangle+}{\sqrt{2}}$,

$|1,-1\rangle^{1u} = \frac{|\beta\beta\beta\alpha\rangle+|\beta\beta\alpha\beta\rangle-|\beta\alpha\beta\beta\rangle-|\alpha\beta\beta\beta\rangle}{2}$

$I^{tot} = 1$, group 3 $\in$ B$_{2u}$  (A10)

$|1,+1\rangle^{2u} = \frac{|\alpha\alpha\alpha\beta\rangle-|\alpha\alpha\beta\alpha\rangle+|\alpha\beta\alpha\alpha\rangle-|\beta\alpha\alpha\alpha\rangle}{2}$,

$|1,0\rangle^{2u} = \frac{|\alpha\beta\alpha\beta\rangle-|\beta\alpha\beta\alpha\rangle}{\sqrt{2}}$,

$|1,-1\rangle^{2u} = \frac{-|\beta\beta\beta\alpha\rangle+|\beta\beta\alpha\beta\rangle-|\beta\alpha\beta\beta\rangle+|\alpha\beta\beta\beta\rangle}{2}$

$I^{tot} = 1$, group 4 $\in$ B$_{3g}$

$|1,+1\rangle^{3g} = \frac{-|\alpha\alpha\alpha\beta\rangle+|\alpha\alpha\beta\alpha\rangle+|\alpha\beta\alpha\alpha\rangle-|\beta\alpha\alpha\alpha\rangle}{2}$,

$|1,0\rangle^{3g} = \frac{|\beta\alpha\alpha\beta\rangle-|\alpha\beta\beta\alpha\rangle}{\sqrt{2}}$,

$|1,-1\rangle^{3g} = \frac{-|\beta\beta\beta\alpha\rangle+|\beta\beta\alpha\beta\rangle+|\beta\alpha\beta\beta\rangle-|\alpha\beta\beta\beta\rangle}{2}$

$I^{tot} = 0$, group 5 $\in$ A$_g$

$|0,0\rangle^{g,s1} = \frac{-\kappa|\beta\beta\alpha\alpha\rangle+|\beta\alpha\beta\alpha\rangle+(\kappa-1)|\beta\alpha\alpha\beta\rangle+(\kappa-1)|\alpha\beta\beta\alpha\rangle+|\alpha\beta\alpha\beta\rangle-\kappa|\alpha\alpha\beta\beta\rangle}{2\sqrt{1-\kappa+\kappa^2}}$,

$I^{tot} = 0$, group 6 $\in A_g$

$$|0,0\rangle^{g,s2} = \frac{(\kappa-2)|\beta\beta\alpha\alpha\rangle-(2\kappa-1)|\beta\alpha\beta\alpha\rangle+(\kappa+1)|\beta\alpha\alpha\beta\rangle+(\kappa+1)|\alpha\beta\beta\alpha\rangle-(2\kappa-1)|\alpha\beta\alpha\beta\rangle+(\kappa-2)|\alpha\alpha\beta\beta\rangle}{2\sqrt{3}\sqrt{1-\kappa+\kappa^2}}.$$

$$\text{with } \kappa = \frac{\sqrt{J_g^2+J_c^2+J_t^2-(J_gJ_c+J_gJ_t+J_cJ_t)}+(J_g-J_c)}{J_g-J_t}$$

It is not as trivial as before to fill the SpinRep line in the character **Table A5** as in the previous cases; this needs some elaboration. However, the first four elements are identical to the one from **Table A4** for group $D_2$.

Let's consider mirror plane $\sigma(yz)$ (**Figure 1A**), that also can be referred to as $\sigma_h$. It does not change the position of atoms (a) and because spin is an axial vector (b) the spin states do not change under the action of $\sigma(yz)$. Hence the corresponding character is 16 (number of spin states). We refer to this operator as parity in the main text, and it changes sign of one coordinate axis (here x).

Now, let's consider two mirror planes $\sigma(xy)$ and $\sigma(xz)$ (**Figure 1A**), also referred to as $\sigma_v$. $\sigma_v$ exchanges two neighbor protons (permutations (12)(34) or (14)(23)) and changes sign of the spin projection (it is not convenient for NMR but using this notation of axis, we assume the projections of the spin states along x axis). So, when two pairs of protons are exchanged and their sign is inverted, then there are only 4 states what do not change under the action of this transformation: $|\alpha\beta\alpha\beta\rangle$, $|\beta\alpha\alpha\beta\rangle$, $|\beta\alpha\beta\alpha\rangle$ and $|\alpha\beta\beta\alpha\rangle$ for the case of (12)(34) permutation with inversion. Analogous 4 states can be written for the other mirror transformation. Hence the two corresponding characters are 4.

And finally, inversion i. It exchanges the protons as (13)(24) (a) and changes sign of the spin projections, meaning that only four states $|\alpha\alpha\beta\beta\rangle$, $|\beta\beta\alpha\alpha\rangle$, $|\beta\alpha\alpha\beta\rangle$ and $|\alpha\beta\beta\alpha\rangle$ will stay the same; hence the corresponding character is 4.

**Table A5.** Table of characters for AA' (AA') four spin-1/2 system ($D_{2h}$ group). This can be obtained as a direct product of $C_i$ (the same as $C_2$) and $D_2$ character groups. See text how to fill SpinRep line.

|   | E | $C_2(z)$ | $C_2(y)$ | $C_2(x)$ | i | $\sigma(xy)$ | $\sigma(xz)$ | $\sigma(yz)$ |
|---|---|---|---|---|---|---|---|---|
| **$A_g$** | +1 | +1 | +1 | +1 | +1 | +1 | +1 | +1 |
| **$B_{1g}$** | +1 | +1 | -1 | -1 | +1 | +1 | -1 | -1 |
| **$B_{2g}$** | +1 | -1 | +1 | -1 | +1 | -1 | +1 | -1 |
| **$B_{3g}$** | +1 | -1 | -1 | +1 | +1 | -1 | -1 | +1 |
| **$A_u$** | +1 | +1 | +1 | +1 | -1 | -1 | -1 | -1 |
| **$B_{1u}$** | +1 | +1 | -1 | -1 | -1 | -1 | +1 | +1 |
| **$B_{2u}$** | +1 | -1 | +1 | -1 | -1 | +1 | -1 | +1 |
| **$B_{3u}$** | +1 | -1 | -1 | +1 | -1 | +1 | +1 | -1 |
| **SpinRep = $7A_g + 3B_{1u}+3B_{1u}+3B_{3g}$** | 16 | 4 | 4 | 4 | 4 | 4 | 4 | 16 |

There are 7 symmetrical ($A_g$ symmetry) states, and three states for each of three B symmetries ($B_{1u}$, $B_{2u}$ and $B_{3g}$). This follows from both eq (A10) and **Table A5**. Therefore, the basis for AA'(AA') four spin-1/2 system of $D_{2h}$ symmetry consists of 4 sets with multiplicity 7, 3, 3 and 3 (Spins rep. = $7A_g + 3B_{1u} + 3E_{2g} + 3E_{3u}$):

$$S_{A_g}^{D_{2h}} = \{|2,+2\rangle^g, |2,+1\rangle^g, |2,0\rangle^g, |2,-1\rangle^g, |2,-2\rangle^g, |0,0\rangle^{g,s1}, |0,0\rangle^{g,s2}\},$$

$$S_{B_{1u}}^{D_{2h}} = \{|1,+1\rangle^{1u}, |1,0\rangle^{1u}, |1,-1\rangle^{1u}\},$$

$$S_{B_{2u}}^{D_{2h}} = \{|1,+1\rangle^{2u}, |1,0\rangle^{2u}, |1,-1\rangle^{2u}\}, \quad (A11)$$

$$S_{B_{3g}}^{D_{2h}} = \{|1,+1\rangle^{3g}, |1,0\rangle^{3g}, |1,-1\rangle^{3g}\}$$

Note that in the case of $C_4$ symmetry considered earlier, six states belong to the $A_{(g)}$ group.

**Appendix B**

**Table B1.** Maximum expected polarization transfer coefficient $\xi_{max}$ for systems without symmetry constraints and for pH$_2$ derived spin order transfer (states $\sigma_{ZZ}^{A,B}$ or $\sigma_{S}^{A,B}$) to the longitudinal polarization of all protons of the same molecule (average polarization).

| Type of the system | Number of spins | $\xi_{max}$ from $\sigma_{ZZ}^{A,B} = \xi\sigma_{P}^{X} + \sigma_{rest}$ | $\xi_{max}$ from $\sigma_{S}^{A,B} = \xi\sigma_{P}^{X} + \sigma_{rest}$ |
|---|---|---|---|
| AB | 2 | ½ | 1 |
| ABC | 3 | ½ | 2/3 |
| ABCD | 4 | 3/8 | 5/8 |
| ABCDE | 5 | 3/8 | 11/20 |
| ABCDEF | 6 | 5/16 | ½ |
| ABCDEFG | 7 | 5/16 | 0.4821 |
| ABCDEFGH | 8 | 0.2734 | 0.4336 |
| ABCDEFGHI | 9 | 0.2734 | 0.4323 |
| ABCDEFGHIJ | 10 | 0.2461 | 0.3906 |
| ABCDEFGHIJK | 11 | 0.2461 | 0.3835 |
| ABCDEFGHIJKL | 12 | 0.2256 | 0.3597 |

**Table B2.** Maximum expected polarization transfer coefficient $\xi_{max}^{SC}$ for systems with symmetry constraints and for pH$_2$ derived spin order transfer (state $\sigma_{S}^{A,B}$) to the longitudinal polarization of all protons of the same molecule (average polarization).

| Type of the system | Number of spins | $\xi_{max}^{SC}$ from $\sigma_{S}^{A,B} = \xi\sigma_{P}^{X} + \sigma_{rest}$ |
|---|---|---|
| A$_2$B | 3 | 1/3 |
| A$_2$BC | 4 | 0.3125 |
| A$_3$B | 4 | 0.25 |
| A$_2$B$_2$ | 4 | 0.1875 |
| A$_3$BC | 5 | 0.175 |
| A$_2$B$_2$C | 5 | 0.175 |
| A$_3$B$_2$ | 5 | 0.1167 |
| A$_3$BCD | 6 | 0.2014 |
| A$_3$B$_2$C | 6 | 0.1424 |
| A$_3$B$_3$ | 6 | 0.1204 |